\documentclass[twocolumn,showpacs,preprintnumbers,amsmath,amssymb,pra]{revtex4}

\usepackage{graphicx}% Include figure files
\usepackage{dcolumn}% Align table columns on decimal point
\usepackage{bm}% bold math
\usepackage{amsmath}
\usepackage{amssymb}
\usepackage{mathrsfs}

\begin{document}

\title{Measuring microwave quantum states: tomogram and moments}% Force line breaks with \\

\author{Sergey N. Filippov}
\email{sergey.filippov@phystech.edu}
\author{Vladimir I. Man'ko}
\email{manko@sci.lebedev.ru} \affiliation{Moscow Institute of
Physics and Technology (State University), 141700 Moscow Region,
Russia} \affiliation{P.~N.~Lebedev Physical Institute, Russian
Academy of Sciences, 119991 Moscow, Russia}

\begin{abstract}
Two measurable characteristics of microwave one-mode photon states
are discussed: a rotated quadrature distribution (tomogram) and
normally/antinormally ordered moments of photon creation and
annihilation operators. Extraction of these characteristics from
amplified microwave signal is presented. Relations between the
tomogram and the moments are found and can be used as a cross
check of experiments. Formalism of the ordered moments is
developed. The state purity and generalized uncertainty relations
are considered in terms of moments. Unitary and non-unitary time
evolution of moments is obtained in the form of a system of linear
differential equations in contrast to partial differential
equations for quasidistributions. Time evolution is specified for
the cases of a harmonic oscillator and a damped harmonic
oscillator which describe noiseless and decoherence processes,
respectively.

\end{abstract}


\maketitle

\section{\label{sec:introduction} Introduction}
Non-classical states of light are of great interest not only from
the viewpoint of academic science but also from the viewpoint of
their promising quantum information applications such as
cryptography, communication, etc. Any quantum experiment consists
of the following procedures: preparation of a desired quantum
state, time evolution of the state (its transformation through a
quantum channel), and a measurement. The latter one is a keystone
to get an insight into quantum world. The result of a measurement
is usually a set of outcomes distributed in accordance with the
quantum nature of the state. Given a numerous number of
identically prepared quantum states and neglecting the memory
effects of channels, one can perform many individual measurements
and gather the statistical information. We refer to the obtained
information as a measurable characteristic of the state. At this
stage one encounters a problem how a measurable characteristic is
related with the quantum state itself, i.e. how to represent the
results of measurements.

A quantum state is usually described by wave function or density
operator~\cite{dirac}. Alternatively, different representations of
the density operator are widely used, e.g., the Wigner
function~\cite{wigner32}. Recently, a probability representation
of quantum states was introduced~\cite{mancini96}. In the
probability representation, quantum states are identified with
fair probability distributions. The probability representation was
introduced in connection with homodyne measuring the photon
quantum states by means of optical tomography. Using the homodyne
detection scheme gives rise to the optical tomogram $w(X,\theta)$
which has a meaning of the probability distribution of a single
rotated quadrature component $\hat{X}_{\theta} =
\hat{q}\cos\theta+\hat{p}\sin\theta$ in the phase space of one
light mode (see, e.g., the review~\cite{leonhardt95}). The aim of
homodyne measurements~\cite{raymer93} was to obtain the Wigner
function identified with the quantum state. Then the reconstructed
Wigner function was used to extract an information on the physical
properties. The measurements of this kind in optical frequency
domain were fulfilled, e.g., in~\cite{mlynek96,solimeno09}, and
the detailed review of the experimental results is presented in
Ref.~\cite{lvovsky09}. Though optical tomogram is a fair
probability distribution, it was interpreted in all the
experimental works as a technical ingredient providing tool to
obtain the Wigner function. The latter one was interpreted as a
`real state'. On the contrary, following the ideas of the
probability
representation~\cite{mancini96,mancini-manko-tombesi-97} (see also
the recent review~\cite{ibort09} and the
paper~\cite{filippov-fss-11}), the optical tomogram and other
kinds of tomographic-probability distributions like symplectic
tomogram~\cite{d'ariano-manko-96} were considered as a primary
object containing complete information on quantum state. In view
of this fact, one does not need reconstructing any
quasidistribution including the Wigner function.  The
reconstruction procedure produces extra inaccuracies related to
the useless elaborating the experimental data by means of Radon
integral transforms. The experiment on direct checking
purity-dependent uncertainty
relations~\cite{dodonov-manko-89,dodonov02} and measuring the
photon state purity and temperature, without reconstruction of the
Wigner function or another quasidistribution, was performed
recently in Ref.~\cite{porzio11}. In this experiment, the optical
tomogram was measured and considered as a primary object
determining the quantum state. Moreover, the above physical
characteristics were expressed in terms of the tomogram, which is
nothing else but directly measurable alternative to the density
operator and the Wigner function.

On the other hand, at optical frequencies there is another
measurable characteristic of the state. Indeed, using the
heterodyne detection scheme gives rise to the Husimi function
$Q(q,p)$ which also contains the full information about the
quantum state~\cite{leonhardt95}. Since both tomogram and Husimi
function are extracted from experimental data, the relation
between them (see, e.g.,~\cite{andreev11}) can be considered as a
cross-check of the experiment accuracy. Closely related to the
Husimi function are ordered moments $\langle (\hat{a}^{\dag})^n
\hat{a}^m \rangle$ and $\langle \hat{a}^k (\hat{a}^{\dag})^l
\rangle$, where $\hat{a}^{\dag}$ and $\hat{a}$ are photon creation
and annihilation operators, respectively. In fact, the moments
also contain the complete information on the state (see, e.g.,
\cite{wunsche90,lee92,herzog96,buzek96}). An optical scheme for
measuring moments is proposed, e.g., in~\cite{shchukin05}. The
moments are complex functions of integers. The measurements give
Husimi function-type characteristics corresponding to a two-mode
state, from which the moments for one-mode state are calculated.
In this sense, the method of moments is not completely direct
measurement in contrast to the optical tomography. Also, the
moments are not a probability distribution as the tomogram is.
Nevertheless, if the moments are known, then all desired
characteristics of the quantum state can be calculated.

The importance of microwaves in up-to-date quantum technologies
can scarcely be overestimated. Using microwave wavelengths
transforms the scale of experiments as compared to standard
optical ones. As a result, at microwaves the highest-quality
superconducting resonators~\cite{kuhr} are built and successfully
used in the microwave version of cavity quantum electrodynamics.
Being applied in one-dimensional resonators together with
superconducting qubits, microwaves have opened an opportunity to
achieve the strongest ever coupling between the electromagnetic
field and an artificial atom (qubit) within compact and integrable
electrical circuits~\cite{blais}. Experimentally realized
couplings of microwaves to transmons~\cite{mallet09} and
mechanical oscillators~\cite{wooley} made them a really
significant tool for further progress of quantum information.

On the other hand, microwave quantum states are of great interest
per se as carriers of quantum information. However, measuring
quantum state of electromagnetic field changes dramatically and
becomes a challenge when microwave radiation is under
investigation. Detection of microwave field (especially itinerant
modes) is complicated by the low efficiency of photodetectors,
although suggestions of high-efficiency microwave photon detectors
are made~\cite{romero09,romero-ps-09}. As a result, it is unlikely
to carry out the photon counting measurements. In order to
register the radiation reliably, amplifiers are widely
used~\cite{castellanos-beltran-08,bergeal10} though they
unavoidably add some noise. In addition, interferometry
experiments in optics extensively utilize beam splitters, whereas
in microwave engineering the role of a beam splitter for a single
mode is played by hybrid junctions or power
dividers~\cite{collin,pozar}. Similarly, a microwave signal is
mixed with a local oscillator microwave field (via a nonlinear
device called mixer) instead of passing a signal mode and a local
oscillator mode through a beam splitter in optics.

In spite of the challenges dealt with microwave radiation, the
measurements of the `optical' tomogram and ordered moments are
feasible and were reported recently
\cite{mallet11,menzel10,mariantoni10,eichler10}. Really, using a
homodyne detection scheme with phase sensitive amplifiers less
noisy than high-electron mobility transistors (e.g., a Josephson
parametric amplifier~\cite{teufel09}) enables, in principle,
measuring tomogram $\tilde{w}(X,\theta)$. This experimentally
accessible distribution $\tilde{w}(X,\theta)$ differs from the
genuine tomogram of the state $w(X,\theta)$ because of the noise
presented. However, the information of the probed state can still
be extracted from the data. An analogue of the 8-port homodyne
detection of optical photons is realized at microwave level by
phase-insensitive amplifiers and an in-phase quadrature (I/Q)
mixer. The I/Q mixer provides two outputs $q$ and $p$ described by
a single envelope function $S=q+ip$. Statistics of experimentally
measurable quantities $q={\rm Re}S$ and $p={\rm Im}S$ allows
constructing the histogram $\tilde{Q}(q,p)$. If there were no
extra noise added, this quasi-probability distribution
$\tilde{Q}(q,p)$ would be an appropriately scaled Husimi function
$Q(q,p)$ of the quantum state. Averaging the Husimi function with
complex function $(S^{\ast})^l S^k = (q-ip)^l(q+ip)^k$ results in
the mean value of anti-normally ordered operator (see,
e.g.,~\cite{schleich}), namely, $\langle \hat{a}^k (\hat{a}
^{\dag})^l \rangle$.

The aim of our paper is to consider both methods of measuring
microwave quantum states, viz., the homodyne tomography and the
measurement of ordered moments. We briefly discuss how to extract
moments from data corresponding to the amplified signal. Using
different experimental schemes, an access to normally and
antinormally ordered moments can be realized. The relations
between the tomogram and the moments are derived, which in case of
`optical' tomogram and normally ordered moments coincide with
relations found in Refs.~\cite{richter96,wunsche96}. This
connections can be utilized to perform cross-check of
measurements. We also suggest a test in the form of inequalities
for experimentally measured moments (both quadrature and
creation-annihilation ones). This test is equivalent to checking
uncertainty relations but avoids unnecessary reconstruction of the
Wigner function or density operator. Finally, we cannot help
considering time evolution of the quantum state in terms of
measurable characteristics. Time evolution in the tomographic
probability representation was considered in
Refs.~\cite{mancini96,ibort09,korennoy11}. In this paper, we fill
this gap for moments. Namely, the Moyal equation~\cite{moyal49}
for the Wigner function is rewritten in terms of moments as well
as an eigenstate problem of Hamiltonian is formulated and
non-unitary evolution of damped electromagnetic field oscillator
is considered for a partial case of generic study of the quantum
oscillator with dissipation~\cite{dodonov2000}. The latter problem
is instructive to clarify decoherence phenomena in microwave
experiments, where the decoherence occurs due to a finite
conductivity of waveguide walls or a lossy dielectric.

The paper is organized as follows.

In Sec.~\ref{sec:moments}, normally and antinormally ordered
moments are extracted from amplified signals and calculated for
examples of Fock, coherent, even/odd
coherent~\cite{dodonov-malkin-manko-74}, and thermal states. In
Sec.~\ref{sec:tomogram} a brief review of symplectic and optical
tomograms is given. In Sec.~\ref{sec:relation-tom-moments},
relations between the tomogram and the moments are derived as well
as purity is expressed in terms of moments. In
Sec.~\ref{sec:uncertainty-relations}, inequalities for moments and
generalized purity-dependent uncertainty relations are suggested
in connection with experiments such as that in
Refs.~\cite{houck07,hofheinz09,eichler10}, where a deterministic
generation of microwave quantum states is demonstrated. In
Sec.~\ref{sec:evolution}, a unitary time evolution and a problem
of Hamiltonian eigenstates are formulated for moments. In
Sec.~\ref{sec:damping}, a particular case of the damped time
evolution of moments is analyzed. Finally, in
Sec.~\ref{sec:conclusions}, conclusions and prospects are given.

\section{\label{sec:moments} Moments}
To anticipate, let us outline a linear parametric amplifier as a
constituent of microwave engineering. The amplifier changes the
quantum state of the signal and idler incident modes by an ${\rm
SU(1,1)}$ transformation~\cite{leonhardt94}, i.e. the Bogoliubov
transformation, which has the following form in the Heisenberg
picture (see, e.g., \cite{caves82,clerk10}):
\begin{eqnarray}
&& \label{signal-amp} \hat{b}_{\rm s} = \sqrt{g} \hat{a} +
\sqrt{g-1}
\hat{h}^{\dag},\\
&& \label{idler-amp} \hat{b}_{\rm i} = \sqrt{g-1} \hat{a}^{\dag} +
\sqrt{g} \hat{h},
\end{eqnarray}
\noindent where $\hat{b}_{\rm s}$ and $\hat{b}_{\rm i}$ are
annihilation operators of the amplified signal and idler modes,
respectively, $\hat{a}$ and $\hat{h}$ are annihilation operators
of the original microwave signal and extra noise modes,
respectively. Depending on the operation of the amplifier, the
output mode can be described by either Eq. (\ref{signal-amp}) or
Eq. (\ref{idler-amp}).

If the experiment exploits a heterodyne detection
scheme~\cite{menzel10,mariantoni10,eichler10}, the measured data
represent nothing else but an appropriately scaled histogram of
the Husimi function $H_{\rm amp}(q,p)$, which differs from the
Husimi function of the original signal $H(q,p)$. However, the
information about the quantum state can be revealed by means of
moments. In fact, averaging $H_{\rm amp}(q,p)$ with complex
function $(S^{\ast})^l S^k = (q-ip)^l(q+ip)^k$ results in the mean
value of anti-normally ordered operator $\hat{b}^k
(\hat{b}^{\dag})^l$~\cite{da-silva-10}, where $\hat{b}$ is
determined by either (\ref{signal-amp}) or (\ref{idler-amp}). For
the sake of brevity, we will consider the case
$\hat{b}=\hat{b}_{\rm s}$ only, which is readily rewritten for the
opposite case $\hat{b}=\hat{b}_{\rm i}$ if needed.

Thus, experimentally accessible moments read
\begin{eqnarray}
\label{mean-moments-amp-a-h} && \langle \hat{b}_{\rm s}^k
(\hat{b}_{\rm s}^{\dag})^l \rangle =
\sum_{i=0}^{k} \sum_{j=0}^{l} \left(%
\begin{array}{c}
  k \\
  i \\
\end{array}%
\right) \left(%
\begin{array}{c}
  l \\
  j \\
\end{array}%
\right) g^{(i+j)/2} \nonumber\\
&& \times (g-1)^{(k+l-i-j)/2} \langle \hat{a}^i (\hat{a}^{\dag})^j
\rangle \langle (\hat{h}^{\dag})^{k-i} \hat{h}^{l-j} \rangle ,
\qquad
\end{eqnarray}
\noindent where the signal mode $a$ and the noise mode $h$ are
assumed to be independent. Once gain $g$ is known, average value
of the normally-ordered noise operators $\langle
(\hat{h}^{\dag})^{m} \hat{h}^{n} \rangle$, $m,n=0,1,\ldots$ can be
measured by analyzing vacuum state $|0\rangle$ of mode $a$ for
which $\langle 0 | \hat{a}^i (\hat{a}^{\dag})^j | 0 \rangle = i!
\delta_{ij}$. Substituting this result in
(\ref{mean-moments-amp-a-h}), we obtain an infinite system of
linear equations of the form
\begin{eqnarray}
\label{mean-moments-amp-h} && \langle {\rm vac} | \hat{b}_{\rm
s}^k (\hat{b}_{\rm s}^{\dag})^l | {\rm vac} \rangle =
\sum_{i=0}^{\min (k,l)}
\frac{k!l!}{i!(k-i)!(l-i)!} \nonumber\\
&& \qquad\qquad \times g^{i} (g-1)^{(k+l)/2-i} \langle
(\hat{h}^{\dag})^{k-i} \hat{h}^{l-i} \rangle. \qquad\qquad
\end{eqnarray}
\noindent Solution of this system is readily found for lower
moments.  If $n=0$ or $m=0$, we obtain
\begin{eqnarray}
&& \langle (\hat{h}^{\dag})^{0} \hat{h}^{n} \rangle = (g-1)^{-n/2}
\langle {\rm vac} |
\hat{b}_{\rm s}^0 (\hat{b}_{\rm s}^{\dag})^n | {\rm vac} \rangle,\\
&& \langle (\hat{h}^{\dag})^{m} \hat{h}^{0} \rangle = (g-1)^{-m/2}
\langle {\rm vac} | \hat{b}_{\rm s}^m (\hat{b}_{\rm s}^{\dag})^0 |
{\rm vac} \rangle. \qquad\quad
\end{eqnarray}
\noindent Similarly, the cases $n=1$ or $m=1$ yield
\begin{eqnarray}
&& \langle (\hat{h}^{\dag})^{1} \hat{h}^{n} \rangle =
(g-1)^{-(1+n)/2} \langle {\rm vac} | \hat{b}_{\rm s}^1
(\hat{b}_{\rm
s}^{\dag})^n | {\rm vac} \rangle \nonumber\\
&&  - n g(g-1)^{-(1+n)/2} \langle {\rm vac} | \hat{b}_{\rm s}^0
(\hat{b}_{\rm s}^{\dag})^{n-1} | {\rm vac} \rangle,\\
&& \langle (\hat{h}^{\dag})^{m} \hat{h}^{1} \rangle =
(g-1)^{-(m+1)/2} \langle {\rm vac} | \hat{b}_{\rm s}^m
(\hat{b}_{\rm
s}^{\dag})^1 | {\rm vac} \rangle \nonumber\\
&&  - m g(g-1)^{-(m+1)/2} \langle {\rm vac} | \hat{b}_{\rm
s}^{m-1} (\hat{b}_{\rm s}^{\dag})^{0} | {\rm vac} \rangle.
\qquad\quad
\end{eqnarray}
\noindent Finally, if moments up to the fourth are of interest,
i.e. $n+m \le 4$ as, e.g., in~\cite{eichler10}, then we add the
missing moment
\begin{eqnarray}
\langle (\hat{h}^{\dag})^{2} \hat{h}^{2} \rangle &=& (g-1)^{-2}
\langle {\rm vac} | \hat{b}_{\rm s}^2 (\hat{b}_{\rm s}^{\dag})^2 |
{\rm vac} \rangle
\nonumber\\
&& - 4g(g-1)^{-2} \langle {\rm vac} | \hat{b}_{\rm s}^1
(\hat{b}_{\rm s}^{\dag})^1 | {\rm vac} \rangle \nonumber\\
&& + 2g^2(g-1)^{-2} \langle {\rm vac} | \hat{b}_{\rm s}^0
(\hat{b}_{\rm s}^{\dag})^0 | {\rm vac} \rangle. \qquad
\end{eqnarray}
Once normally ordered moments of noise mode are calculated, the
antinormally ordered moments of the microwave radiation mode are
found from experimental data by solving the linear system
(\ref{mean-moments-amp-a-h}) with respect to $\langle \hat{a}^i
(\hat{a}^{\dag})^j \rangle$. In the same way, the choice
$\hat{b}=\hat{b}_{\rm i}$ enables calculation of the antinormally
ordered moments of noise mode and normally ordered moments of the
original microwave mode. Relations between differently ordered
moments are given in Ref.~\cite{wunsche99}.

Since ordered moments are extracted from experimental data, it is
reasonable to compare them with theoretical values. Let us
consider examples of quantum one-mode states and calculate the
corresponding normally and antinormally ordered moments.

As far as Fock states $|N\rangle$ are concerned, we have
\begin{eqnarray}
&& \label{Fock-norm} \langle N | (\hat{a}^{\dag})^n \hat{a}^m | N \rangle = \left\{%
\begin{array}{cll}
  \dfrac{N!}{(N-n)!} & & {\rm if~} n = m \le N , \\
  0 & & {\rm otherwise}; \\
\end{array}%
\right.\qquad\\
&& \label{Fock-antinorm} \langle N | \hat{a}^k (\hat{a}^{\dag})^l
| N \rangle = \frac{(N+k)!}{N!} \delta_{kl}.
\end{eqnarray}

Among mixed states, we point out a thermal state given by the
unitless temperature $T$. Moments are obviously related with the
average photon number and read
\begin{eqnarray}
&& \label{thermal-norm} \langle (\hat{a}^{\dag})^n \hat{a}^m
\rangle_{\rm thermal} =
\frac{n!\delta_{nm}}{\left( e^{1/T}-1 \right)^n},\\
&& \langle \hat{a}^k (\hat{a}^{\dag})^l \rangle_{\rm thermal} =
\frac{k!\delta_{kl}}{\left( 1-e^{-1/T} \right)^k}.
\end{eqnarray}

Classical-like states are represented by a family of coherent
states $|\alpha\rangle$, i.e. eigenstates of the annihilation
operator $\hat{a}|\alpha\rangle = \alpha |\alpha\rangle$. These
states can be generated, e.g., by masers, and represent an
electromagnetic field of a local oscillator if $|\alpha| \gg 1$.
The ordered moments are
\begin{eqnarray}
&& \!\!\!\!\!\!\! \langle \alpha | (\hat{a}^{\dag})^n \hat{a}^m |
\alpha \rangle =
(\alpha^{\ast})^{n} \alpha^m,\\
&& \!\!\!\!\!\!\! \langle \alpha | \hat{a}^k (\hat{a}^{\dag})^l |
\alpha \rangle =
\sum_{p=0}^{\min(k,l)} \left(%
\begin{array}{c}
  k \\
  p \\
\end{array}%
\right) \left(%
\begin{array}{c}
  l \\
  p \\
\end{array}%
\right) p! \alpha^{k-p} (\alpha^{\ast})^{l-p}. \qquad
\end{eqnarray}

Other examples of nonclassical pure states of light are even/odd
coherent states $|\pm \rangle = \mathcal{N}_{\pm} \left(
|\alpha\rangle \pm |-\alpha\rangle \right)$ with
$\mathcal{N}_{\pm} = \big[2\big(1 \pm \exp (-2|\alpha|^2)
\big)\big]^{-1/2}$, which were introduced in
Ref.~\cite{dodonov-malkin-manko-74}. Transformation of these
states after passing a linear amplifier is considered in
Ref.~\cite{al-kader-01}. Here, we calculate all the ordered
moments and then explore their non-unitary evolution in
Sec.~\ref{sec:damping}. We have
\begin{eqnarray}
&& \label{odd-norm} \!\!\!\!\!\!\! \langle \pm |
(\hat{a}^{\dag})^n \hat{a}^m | \pm \rangle =
\mathcal{N}_{\pm}^{\,2} (\alpha^{\ast})^n \alpha^m
\nonumber\\
&& \!\!\!\!\!\!\! \times \left[ 1 + (-1)^{n+m} \pm
e^{-2|\alpha|^2}\big( (-1)^n +
(-1)^m \big) \right],\\
&& \!\!\!\!\!\!\! \langle \pm | \hat{a}^k (\hat{a}^{\dag})^l | \pm
\rangle =
\mathcal{N}_{\pm}^{\,2} \!\!\! \sum_{p=0}^{\min(k,l)} \!\!\! \left(%
\begin{array}{c}
  k \\
  p \\
\end{array}%
\right) \!\! \left(%
\begin{array}{c}
  l \\
  p \\
\end{array}%
\right) \! p! \ \alpha^{k-p}
(\alpha^{\ast})^{l-p} \nonumber\\
&& \!\!\!\!\!\!\! \times \left[ 1 \! + \! (-1)^{k+l-2p} \pm
e^{-2|\alpha|^2}\big( (-1)^{k-p} \! + \! (-1)^{l-p} \big) \right].
\qquad
\end{eqnarray}

\section{\label{sec:tomogram} Tomogram}
As it is stated in Sec.~\ref{sec:introduction}, apart from the
density operator $\hat{\rho}$, the state of one-mode photon state
can alternatively be described by the symplectic
tomogram~\cite{d'ariano-manko-96}
\begin{equation}
w_{\rm s}(X,\mu,\nu) = {\rm Tr} \big[ \hat{\rho} \delta(X-\mu
\hat{q} - \nu \hat{p}) \big],
\end{equation}
\noindent where $X$ is the real argument, $\mu$ and $\nu$ are real
parameters, $\hat{q}$ and $\hat{p}$ are quadrature operators. In
what follows, we will be interested in the measurable
characteristic, viz., the so-called `optical' tomogram
$w(X,\theta) = w_{\rm s}(X,\mu = \cos\theta,\nu=\sin\theta)$,
where $\theta\in[0,2\pi]$ is the phase of the local oscillator.
Here, we use the adjective `optical' as a conventional one,
although microwave quantum states are under study. Unless
otherwise stated, we will refer to the `optical' tomogram as
tomogram simply.

Analysis of the microwave quantum states employs linear
amplifiers, so the tomogram $w_{\rm amp}(X,\theta)$ of the
amplified signal is only accessible. Our goal is to obtain the
relation between the tomogram of amplified quantum state and the
tomogram of the original one. As shown in Ref.~\cite{kim97}, the
amplification (\ref{signal-amp}) is represented by a convolution
of the signal and idler fields in phase space. The Wigner function
of the amplified signal reads~\cite{kim97}
\begin{equation}
\label{W-convolution} W_{\rm amp}(\alpha) = \frac{1}{g} \int d^2
\beta ~ W\left( \frac{\alpha-\sqrt{g-1}\beta}{\sqrt{g}} \right)
W_{\rm noise} (\beta),
\end{equation}
\noindent where $\alpha=(q+ip)/\sqrt{2}$. Assuming $g \gg 1$, Eq.
(\ref{W-convolution}) can be readily simplified. Integrating such
a simplified Wigner function with delta-function
$\delta(X-q\cos\theta-p\sin\theta)$ yields the convolution
expression for tomograms
\begin{eqnarray}
\label{w-amp} w_{\rm amp}(X,\theta) & = & \frac{1}{\sqrt{g}} \int
d^2\beta ~ w\left( \frac{X}{\sqrt{g}} - \frac{\beta e^{i\theta} +
\beta^{\ast}e^{-i\theta}}{\sqrt{2}},\theta\right) \nonumber\\
&& \qquad \times W_{\rm noise}(\beta).
\end{eqnarray}

Associating $w_{\rm amp}(X,\theta)$ with measured data, it is
possible, in general, to get insight into the state tomogram
$w(X,\theta)$ itself by fulfilling the deconvolution of
(\ref{w-amp}), provided the thorough knowledge of the noise state.
The noise mode is usually assumed to be in thermal state with an
experimentally controllable temperature $T$~\cite{mariantoni10}.
If this is the case, then
\begin{eqnarray}
\label{tom-amp-tom-noise-temper}
\!\!\!\!\! && w_{\rm amp}(X,\theta) \nonumber\\
\!\!\!\!\! && = \frac{1}{\sqrt{2\pi \sigma^2 g}}
\int\limits_{-\infty}^{+\infty} w\left( \frac{X}{\sqrt{g}} - Y,
\theta \right) \exp \left[ - \frac{Y^2}{2\sigma^2} \right] dY,
\qquad
\end{eqnarray}
\noindent where Gaussian variance $\sigma$ is expressed in terms
of the unitless temperature $T$ via formula $\sigma =
\sqrt{\frac{1}{2} \coth\frac{1}{2T}}$.

While processing the experimental data, it is convenient to deal
with tomographic moments $\langle X_{\theta}^{r} \rangle =
\int_{-\infty}^{+\infty} X^r w(X,\theta) d \theta$. These moments
are known to contain the full information about the quantum state.
Using (\ref{tom-amp-tom-noise-temper}), we get the connection
between tomographic moments of the amplified signal $\langle
X_{\theta}^{r} \rangle_{\rm amp}$ and tomographic moments of the
original microwave photon state $\langle X_{\theta}^{i} \rangle$:
\begin{eqnarray}
\label{tomographic-moments-amp} \langle X_{\theta}^{r}
\rangle_{\rm amp} = \sqrt{g^r}
\sum\limits_{l=0}^{[r/2]} \left(%
\begin{array}{c}
  r \\
  2l \\
\end{array}%
\right) (2l-1)!! \langle X_{\theta}^{r-2l} \rangle \sigma^{2l},
\qquad
\end{eqnarray}
\noindent where $(2l-1)!! = 1 \cdot 3 \cdot 5 \cdot \ldots \cdot
(2l-1)$ and $(-1)!! = 1$. All the tomographic moments $\langle
X_{\theta}^{i} \rangle$ up to a desired order $R$ can be readily
obtained from experimentally accessible tomographic moments
$\langle X_{\theta}^{r} \rangle_{\rm amp}$, $r = 0, 1, \ldots, R$,
by inversion of formula (\ref{tomographic-moments-amp}). After
that, the tomographic probability-distribution function can be
also expressed through tomographic moments.

\section{\label{sec:relation-tom-moments} Relation between tomograms and moments}
Let us now derive relations between measurable moments and
measurable tomographic distributions. These relations are to be
used as a cross check of two approaches to probing microwave
quantum states.

Using the formalism of characteristic functions $\langle
e^{\lambda \hat{a}^{\dag}} e^{-\lambda^{\ast} \hat{a}} \rangle$
and $\langle e^{-\lambda^{\ast} \hat{a}} e^{\lambda
\hat{a}^{\dag}} \rangle$, it is not hard to prove that the Wigner
function is expressed through normally-ordered and
antinormally-ordered moments as follows:
\begin{eqnarray}
\label{W-norm} W(\alpha) &=& \sum_{n,m} \frac{\langle
(\hat{a}^{\dag})^n \hat{a}^m \rangle}{\pi^2 n! m!} \int d^2
\lambda ~ \lambda^n
(-\lambda^{\ast})^m \nonumber\\
&& \times \exp
\left[-\frac{1}{2}|\lambda|^2+\alpha\lambda^{\ast}-\alpha^{\ast}\lambda\right],\\
\label{W-antinorm} W(\alpha) &=& \int d^2 \lambda \left(
\sum_{k,l} \frac{\langle \hat{a}^k (\hat{a}^{\dag})^l \rangle
(-\lambda^{\ast})^k
\lambda^l}{\pi^2 k! l!} \right) \nonumber\\
&& \times \exp \left[
\frac{1}{2}|\lambda|^2+\alpha\lambda^{\ast}-\alpha^{\ast}\lambda
\right].
\end{eqnarray}

It is worth noting that the expression (\ref{W-norm}) implies
integration over complex plane and then summation on
$n,m=0,1,\ldots$, whereas formula (\ref{W-antinorm}) avoids
singularities if the summation is done before the integration. As
it is shown later, this peculiarity leads to simpler expressions
for normally ordered moments than for antinormally ordered ones.
In Ref.~\cite{bednorz11} a discussion is presented how to reveal
negativity of the Wigner function (\ref{W-norm}) by the analysis
of higher-order moments.

Our goal is the relation between the `optical' tomogram and the
moments but at first we exploit the known mapping of the Wigner
function onto symplectic tomogram (see,
e.g.,~\cite{d'ariano-manko-96}) and obtain after simplification

\begin{eqnarray}
\label{symplectic-norm} && \!\!\!\!\!\!\! w(X,\mu,\nu) =
\frac{1}{\sqrt{\pi}} \exp\left[-\frac{X^2}{\mu^2+\nu^2}\right]
\sum\limits_{n,m} \frac{\langle (\hat{a}^{\dag})^{n} \hat{a}^{m}
 \rangle}{n!m!} \qquad \nonumber\\
&& \!\!\!\!\!\!\! \times \frac{(\mu+i\nu)^n
(\mu-i\nu)^m}{\sqrt{2^{n+m}(\mu^2+\nu^2)^{(n+m+1)}}} ~
H_{n+m}\left( \frac{X}{\sqrt{\mu^2+\nu^2}} \right),\quad\\
\label{symplectic-antinorm} && \!\!\!\!\!\!\! w(X,\mu,\nu) =
\frac{1}{2\pi} \int d\xi ~ \exp\left[ \frac{\xi^2}{4}(\mu^2+\nu^2)
- i \xi X \right] \ \
\nonumber\\
&& \!\!\!\!\!\!\! \times \left( \sum\limits_{k,l} \frac{\langle
\hat{a}^{k} (\hat{a}^{\dag})^{l}
 \rangle}{k!l!} \left(\frac{i\xi}{\sqrt{2}}\right)^{k+l} (\mu-i\nu)^k
(\mu+i\nu)^l \right),\ \
\end{eqnarray}
\noindent where $H_{N}(X)$ is the Hermite polynomial of degree
$N$. Substituting $\cos\theta$ for $\mu$ and $\sin\theta$ for
$\nu$ in (\ref{symplectic-norm})--(\ref{symplectic-antinorm}), the
`optical' tomogram $w(X,\theta)$ is expressed in terms of ordered
moments
\begin{eqnarray}
&& \label{tom-norm-moments} \!\!\!\!\!\!\!\!\!\!\!\!\! w(X,\theta)
= \frac{e^{-X^2}}{\sqrt{\pi}} \sum\limits_{n,m} \frac{\langle
(\hat{a}^{\dag})^{n} \hat{a}^{m}
 \rangle e^{i(n-m)\theta}}{\sqrt{2^{n+m}}n!m!} ~
H_{n+m}(X)\\
&& \label{tom-antinorm-moments} \!\!\!\!\!\!\!\!\!\!\!\!\! = \!
\int \! \frac{d\xi}{2\pi} ~ e^{\xi^2/4 - i \xi X } \! \bigg( \!
\sum\limits_{k,l} \frac{\langle \hat{a}^{k} (\hat{a}^{\dag})^{l}
 \rangle}{k!l!} \left(\tfrac{i\xi}{\sqrt{2}}\right)^{\! k+l} \!\!\! e^{i(l-k)\theta}  \bigg).\ \
\end{eqnarray}

Since both tomogram and moments are measurable in microwave
experiments, Eq. (\ref{tom-norm-moments}) is the basis for a
cross-check of homodyne and heterodyne detection schemes.

Let us now address the problem to invert formulas
(\ref{tom-norm-moments})--(\ref{tom-antinorm-moments}). Using
orthogonality of Hermite polynomials and orthogonality of
trigonometric functions, we have the following inverse relation:
\begin{eqnarray}
\label{norm-mom-tomogram}\langle (\hat{a}^{\dag})^{n} \hat{a}^{m}
\rangle &=& \frac{n!m!}{2\pi\sqrt{2^{n+m}}(n+m)!}
\int\limits_{0}^{2\pi}
d\theta \int\limits_{-\infty}^{+\infty} dX ~ w(X,\theta) \nonumber\\
&& \times  e^{i(m-n)\theta}
 H_{n+m}(X) \\
&& \label{norm-mom-tom-mom}
\!\!\!\!\!\!\!\!\!\!\!\!\!\!\!\!\!\!\!\!\!\!\!\!\!\!\!\!\!\!\!\!\!\!\!\!
= \frac{n!m!}{2\pi\sqrt{2^{n+m}}(n+m)!} \int\limits_{0}^{2\pi}
d\theta ~ e^{i(m-n)\theta}
 \langle H_{n+m}( X_{\theta} ) \rangle,
\end{eqnarray}
\noindent where $\langle H_{n+m}( X_{\theta} ) \rangle$ depends on
$\theta$ and is obtained from polynomial $H_{n+m}( X )$ by
replacing $X^r \rightarrow \langle X_{\theta}^r \rangle$ for all
$r=0,1,\ldots,n+m$. Thus, formula (\ref{norm-mom-tom-mom})
establishes a relation between normally ordered moments and
tomographic moments. We must note that the relation
(\ref{norm-mom-tomogram}) was found previously in the
papers~\cite{richter96,wunsche96}.

As far as antinormally ordered moments are concerned, a direct use
of Eq. (\ref{tom-antinorm-moments}) is sophisticated so we resort
to the connection between antinormally and normally ordered
moments and exploit result of Eqs.
(\ref{norm-mom-tomogram})--(\ref{norm-mom-tom-mom}) to obtain

\begin{eqnarray}
\label{antinorm-mom-tomogram} \langle \hat{a}^{k}
(\hat{a}^{\dag})^{l} \rangle &=& \frac{k!l!}{2\pi\sqrt{2^{k+l}}}
\int\limits_{0}^{2\pi} d\theta
\int\limits_{-\infty}^{+\infty} dX ~ w(X,\theta) ~ e^{i(k-l)\theta} \nonumber\\
&& \times  \sum_{p=0}^{\min(k,l)} \frac{2^p
H_{k+l-2p}(X)}{p!(k+l-2p)!} \\
&& \label{antinorm-mom-tom-mom}
\!\!\!\!\!\!\!\!\!\!\!\!\!\!\!\!\!\!\!\!\!\!\!\!\!\!\!\!\!\!\!\!\!\!\!\!
= \frac{k!l!}{2\pi\sqrt{2^{k+l}}} \int\limits_{0}^{2\pi} d\theta ~
e^{i(k-l)\theta} \!\! \sum_{p=0}^{\min(k,l)} \! \frac{2^p \langle
H_{k+l-2p}(X_{\theta}) \rangle }{p!(k+l-2p)!} .
\end{eqnarray}

In order to perform cross check of the experimental data (tomogram
vs. moments), one can also compare measured tomographic moments
$\langle X_{\theta}^r \rangle$ with tomographic moments predicted
by the measurement of normally ordered moments $\langle
(\hat{a}^{\dag})^n \hat{a}^m \rangle$, namely,
\begin{eqnarray}
\label{tom-mom-norm-mom}
\langle X_{\theta}^{r} \rangle = \!\!\!\!\!\! \sum\limits_{\substack{n+m \le r \\
r-n-m ~ {\rm is~even}}} \!\!\!\!\!\!
\frac{r!\sqrt{2^{n+m-2r}}}{n!m!\left(\frac{r-n-m}{2}\right)!}
\langle (\hat{a}^{\dag})^n \hat{a}^m \rangle
e^{i(n-m)\theta}.\quad
\end{eqnarray}
\noindent The latter formula is obtained by combining a definition
of the tomographic moment $\langle X_{\theta}^r \rangle$, Eq.
(\ref{tom-norm-moments}), and the following integral
\begin{eqnarray}
&& \int\limits_{-\infty}^{+\infty} X^r H_{N}(X) e^{-X^2} dX \nonumber\\
&& = \left\{%
\begin{array}{lll}
  0 & & {\rm if ~} r-N {\rm ~ is ~ odd ~ or ~} N>r,\\
  \cfrac{\sqrt{\pi}r!}{2^{r-N} \left( \frac{r-N}{2} \right)!} & & {\rm otherwise},\\
\end{array}%
\right. \qquad
\end{eqnarray}
\noindent which can also be written in the form of two-dimensional
Hermite polynomial~\cite{dodonov-manko-jpa-94}.

\subsection{Purity in terms of moments}
In this subsection, we attack the problem how to determine purity
of the state without unnecessary reconstruction of the density
operator but dealing with measured normally ordered moments. Just
as the measurable tomogram is used for fast and reliable
calculation of the purity parameter~\cite{porzio11}, here we
derive the purity in terms of moments only.

To start with, the overlap ${\rm Tr}[\hat{\rho}_1 \hat{\rho}_2]$
between two density operators $\hat{\rho}_1$ and $\hat{\rho}_2$
equals to the overlap of the corresponding Wigner functions. In
view of the relation (\ref{W-norm}), it is readily seen that
\begin{eqnarray}
{\rm Tr}[\hat{\rho}_1 \hat{\rho}_2] &=&  \sum_{n,m,k,l}
\frac{(-1)^{m+k} (n+k)!}{n!m!k!l!} \delta_{n+k,m+l} \nonumber\\
&& \qquad \qquad \times \langle (\hat{a}^{\dag})^n \hat{a}^m
\rangle_1 \langle (\hat{a}^{\dag})^k \hat{a}^l \rangle_2.
\end{eqnarray}

For example, if one is interested in how close the prepared state
is to the vacuum one, it is enough to calculate $ \langle 0 |
\hat{\rho} | 0 \rangle = \sum_{k} (-1)^k (k!)^{-1} \langle
(\hat{a}^{\dag})^k \hat{a}^k \rangle$.

Finally, the state is thoroughly described by moments $\langle
(\hat{a}^{\dag})^n \hat{a}^m \rangle$ and its purity $\tilde{\pi}$
is
\begin{equation}
\label{purity} \tilde{\pi}  = \!\! \sum_{n,m,k,l} \!\!
\frac{(-1)^{m+k} (n+k)!}{n!m!k!l!} ~\delta_{n+k,m+l} \langle
(\hat{a}^{\dag})^n \hat{a}^m \rangle \langle (\hat{a}^{\dag})^k
\hat{a}^l \rangle.
\end{equation}

Purity of the thermal state (\ref{thermal-norm}) is
$\tilde{\pi}_{\rm thermal} = \tanh(1/2T)$ so the effective
temperature $T_{\rm eff}$ of the electromagnetic field can be
probed directly via measured moments as $T_{\rm eff} = 1 /(2 ~{\rm
arctanh} \tilde{\pi})$.

\section{\label{sec:uncertainty-relations} Uncertainty relations}
The role of uncertainty relations in quantum mechanics can
scarcely be overestimated. Therefore, a direct probing of
uncertainty relations in experiments is of great interest. The
information in amplified microwave signals is obscured by the
noise. However, as it is shown in Sec.~\ref{sec:moments}, the
ordered moments of the original microwave mode can still be
extracted from the data. The question arises itself whether such
extracted moments satisfy the uncertainty relations. The violation
of these inequalities would indicate the incorrectness of data
processing or measurement of the data.

Let us start with a conventional Schr\"{o}dinger-Robertson
inequality $\sigma_{qq}\sigma_{pp} - \sigma_{qp}^2 \ge 1/4$.
Written in the form of tomographic moments, i.e. $\langle (\Delta
X_{\theta})^2 \rangle = \langle X_{\theta}^2 \rangle - \langle
X_{\theta} \rangle^2$, this inequality takes the
form~\cite{mmsv09}
\begin{eqnarray}
&& \langle (\Delta X_{\theta})^2 \rangle \langle (\Delta
X_{\theta+\pi/2})^2 \rangle - \bigg[ \langle (\Delta
X_{\theta+\pi/4})^2 \rangle \nonumber\\
&& \qquad\quad - \frac{1}{2} \Big( \langle (\Delta X_{\theta})^2
\rangle + \langle (\Delta X_{\theta+\pi/2})^2 \rangle \Big)
\bigg]^{2} \ge \frac{1}{4}\qquad\quad
\end{eqnarray}
\noindent and is to be satisfied for any angle
$\theta\in[0,2\pi]$. Using the established relation between
tomographic moments and normally ordered moments
(\ref{tom-mom-norm-mom}), we have $\langle X_{\theta} \rangle =
\left(\langle \hat{a}^{\dag} \rangle e^{i\theta} + \langle \hat{a}
\rangle e^{-i\theta} \right)/\sqrt{2}$ and $\langle X_{\theta}^2
\rangle = 1/2 + \langle \hat{a}^{\dag} \hat{a} \rangle +
\left(\langle (\hat{a}^{\dag})^2 \rangle e^{i2\theta} + \langle
\hat{a}^2 \rangle e^{-i2\theta}\right)/2$. Finally, we obtain the
uncertainty relation in terms of moments
\begin{eqnarray}
\label{UR-simple} && \left( \langle \hat{a}^{\dag} \hat{a} \rangle
- \langle \hat{a}^{\dag} \rangle \langle \hat{a} \rangle \right) +
\left( \langle \hat{a}^{\dag} \hat{a} \rangle - \langle
\hat{a}^{\dag}
\rangle \langle \hat{a} \rangle \right)^2 \nonumber\\
&& \qquad - \left(\langle (\hat{a}^{\dag})^2 \rangle - \langle
\hat{a}^{\dag} \rangle^2 \right) \left( \langle \hat{a}^2 \rangle
- \langle \hat{a} \rangle^2 \right) ~ \ge ~ 0, \qquad\qquad
\end{eqnarray}
\noindent which turns out to be independent on $\theta$.

The inequality (\ref{UR-simple}) can be made stronger if we take
into account the purity of the state. Indeed, we
have~\cite{dodonov-manko-89} $\sigma_{qq}\sigma_{pp} -
\sigma_{qp}^2 \ge \Phi^2(\tilde{\pi})/4$, where
$\Phi(\tilde{\pi})\approx
(4+\sqrt{16+9\tilde{\pi}^2})/9\tilde{\pi}$ within the accuracy of
1\%. Consequently, purity-dependent uncertainty relation in terms
of normally ordered moments is
\begin{eqnarray}
\label{UR-complex} && \!\!\!\!\!\! \left( \langle \hat{a}^{\dag}
\hat{a} \rangle - \langle \hat{a}^{\dag} \rangle \langle \hat{a}
\rangle \right) + \left( \langle \hat{a}^{\dag} \hat{a} \rangle -
\langle \hat{a}^{\dag}
\rangle \langle \hat{a} \rangle \right)^2 \nonumber\\
&& \!\!\!\!\!\! - \left(\langle (\hat{a}^{\dag})^2 \rangle -
\langle \hat{a}^{\dag} \rangle^2 \right) \left( \langle \hat{a}^2
\rangle - \langle \hat{a} \rangle^2 \right)  \ge
\left(\Phi^2(\tilde{\pi})-1\right)/4, \qquad
\end{eqnarray}
\noindent where the purity $\tilde{\pi}$ is to be calculated via
moments in accordance with Eq. (\ref{purity}). Thus, we have
formulated a self-consistent problem of checking purity-dependent
uncertainty relations by using measurable moments only.

Another drawback of inequality (\ref{UR-simple}) is that it
exploits the lowest moments only, i.e. $\langle (\hat{a}^{\dag})^n
\hat{a}^m \rangle$ with $n+m \le 2$. Let us derive uncertainty
relations in terms of moments such that they involve moments up to
a desired order. In fact, suppose the operator $\hat{F} = z_0
\openone + \sum_{n} (y_n \hat{a}^n +z_n (\hat{a}^{\dag})^n )$ with
arbitrary complex numbers $z_0, y_n, z_n$. Then the mean value
$\langle \hat{F}^{\dag} \hat{F} \rangle \ge 0$ for all $z_0, y_n,
z_n \in \mathbb{C}$. This immediately implies that the quadratic
form is positive-semidefinite. According to Sylvester's criterion,
all leading principal minors of the corresponding matrix are
non-negative. For instance, we readily obtain the inequalities on
moments up to the fourth order:
\begin{equation}
\label{matrix-inequality}
\left(%
\begin{array}{ccccc}
  \langle \openone \rangle & \langle \hat{a} \rangle & \langle \hat{a}^{\dag} \rangle & \langle \hat{a}^2 \rangle & \langle (\hat{a}^{\dag})^2 \rangle \\
  \langle \hat{a}^{\dag} \rangle & \langle \hat{a}^{\dag}\hat{a} \rangle & \langle (\hat{a}^{\dag})^2 \rangle & \langle \hat{a}^{\dag}\hat{a}^2 \rangle & \langle (\hat{a}^{\dag})^3 \rangle \\
  \langle \hat{a} \rangle & \langle \hat{a}^2 \rangle & \langle \hat{a}\hat{a}^{\dag} \rangle & \langle \hat{a}^3 \rangle & \langle \hat{a}(\hat{a}^{\dag})^2 \rangle \\
  \langle (\hat{a}^{\dag})^2 \rangle & \langle (\hat{a}^{\dag})^2\hat{a} \rangle & \langle (\hat{a}^{\dag})^3 \rangle & \langle (\hat{a}^{\dag})^2 \hat{a}^2 \rangle & \langle (\hat{a}^{\dag})^4 \rangle \\
  \langle \hat{a}^2 \rangle & \langle \hat{a}^3 \rangle & \langle \hat{a}^2\hat{a}^{\dag} \rangle & \langle \hat{a}^4 \rangle & \langle \hat{a}^2(\hat{a}^{\dag})^2 \rangle \\
\end{array}%
\right) \ge 0.
\end{equation}
\noindent If normally ordered moments are measured, then one
should replace $ \langle \hat{a}\hat{a}^{\dag} \rangle = \langle
\hat{a}^{\dag}\hat{a} \rangle + 1$, $\langle
\hat{a}^2\hat{a}^{\dag} \rangle = \langle \hat{a}^{\dag}\hat{a}^2
\rangle + 2 \langle \hat{a} \rangle$, $ \langle
\hat{a}(\hat{a}^{\dag})^2 \rangle = \langle
(\hat{a}^{\dag})^2\hat{a} \rangle + 2 \langle \hat{a}^{\dag}
\rangle$, $\langle \hat{a}^2(\hat{a}^{\dag})^2 \rangle = \langle
(\hat{a}^{\dag})^2\hat{a}^2 \rangle + 4 \langle \hat{a}^{\dag}
\hat{a} \rangle + 2$. Otherwise, i.e. if antinormally ordered
moments are experimentally determined, then one should replace
$\langle \hat{a}^{\dag}\hat{a} \rangle = \langle
\hat{a}\hat{a}^{\dag} \rangle - 1$, $ \langle
\hat{a}^{\dag}\hat{a}^2 \rangle = \langle \hat{a}^2\hat{a}^{\dag}
\rangle - 2 \langle \hat{a} \rangle$, $ \langle
(\hat{a}^{\dag})^2\hat{a} \rangle = \langle
\hat{a}(\hat{a}^{\dag})^2 \rangle - 2 \langle \hat{a}^{\dag}
\rangle$, $ \langle (\hat{a}^{\dag})^2\hat{a}^2 \rangle = \langle
\hat{a}^2(\hat{a}^{\dag})^2 \rangle - 4 \langle \hat{a}
\hat{a}^{\dag} \rangle + 2$. Note that formula (\ref{UR-simple})
is nothing else but a condition on nonnegativity of the second
principal minor of the matrix (\ref{matrix-inequality}).

\section{\label{sec:evolution} Unitary evolution and eigenstates}
In this section, we follow ideas of the seminal
paper~\cite{moyal49}. Normally and antinormally ordered moments
are functions of non-commuting operators $\hat{a}$ and
$\hat{a}^{\dag}$ and can be considered as specific `phase-space
quasidistributions'. Now we are aimed at deriving the laws which
govern the transformation with time of these `phase-space
quasidistributions'. In other words, the problem is to find time
evolution equations for ordered moments. If this problem is
solved, then such equations can be used as an alternative to the
Schr\"{o}dinger equation for the wave function, the von Neumann
equation for the density operator, and the Moyal equation for the
Wigner function. The crucial point is that ordered moments are
experimentally measurable characteristics of microwave quantum
states in contrast to wave functions, density operators, and
Wigner functions.

We start with correspondence rules between the operators acting on
the Wigner function and the operators acting on the ordered
moments. Applying operators $q$, $p$, $\partial/\partial q$, and
$\partial/\partial p$ to the left-hand side of Eq. (\ref{W-norm}),
we obtain
\begin{eqnarray}
&& \label{rules-norm-1} \!\!\!\!\!\!\!\!\!\!\!\!\!\!\!\!\!\!\!\!
\frac{\partial}{\partial q} \leftrightarrow -\tfrac{1}{\sqrt{2}}
\big( n \hat{\Delta}_{-1,0}^{({\rm n})} + m \hat{\Delta}_{0,-1}^{({\rm n})} \big),\\
&& \!\!\!\!\!\!\!\!\!\!\!\!\!\!\!\!\!\!\!\!
\frac{\partial}{\partial p} \leftrightarrow \tfrac{i}{\sqrt{2}}
\big( n \hat{\Delta}_{-1,0}^{({\rm n})} - m \hat{\Delta}_{0,-1}^{({\rm n})} \big),\\
&& \!\!\!\!\!\!\!\!\!\!\!\!\!\!\!\!\!\!\!\! q  \leftrightarrow
\tfrac{1}{\sqrt{2}} \big( \hat{\Delta}_{+1,0}^{({\rm n})} \! + \!
\hat{\Delta}_{0,+1}^{({\rm n})} \big) \! + \! \tfrac{1}{2\sqrt{2}}
\big( n \hat{\Delta}_{-1,0}^{({\rm n})} \! + \! m \hat{\Delta}_{0,-1}^{({\rm n})} \big),\\
&& \label{rules-norm-4} \!\!\!\!\!\!\!\!\!\!\!\!\!\!\!\!\!\!\!\! p
\leftrightarrow \tfrac{i}{\sqrt{2}} \big(
\hat{\Delta}_{+1,0}^{({\rm n})} \! - \! \hat{\Delta}_{0,+1}^{({\rm
n})} \big) \! - \! \tfrac{i}{2\sqrt{2}} \big( n
\hat{\Delta}_{-1,0}^{({\rm n})} \! - \! m
\hat{\Delta}_{0,-1}^{({\rm n})} \big),
\end{eqnarray}
\noindent where a displacement operator $\hat{\Delta}_{i,j}^{({\rm
n})}$ for the normally ordered moments is introduced as follows:
\begin{equation}
\hat{\Delta}_{i,j}^{({\rm n})} \langle (\hat{a}^{\dag})^n
\hat{a}^m \rangle := \langle (\hat{a}^{\dag})^{n+i} \hat{a}^{m+j}
\rangle.
\end{equation}

Arguing as above and employing Eq. (\ref{W-antinorm}), the
correspondence relations for antinormally ordered moments are
found
\begin{eqnarray}
&& \label{rules-antinorm-1}
\!\!\!\!\!\!\!\!\!\!\!\!\!\!\!\!\!\!\!\! \frac{\partial}{\partial
q} \leftrightarrow  -\tfrac{1}{\sqrt{2}}
\big( k \hat{\Delta}_{-1,0}^{({\rm a})} + l \hat{\Delta}_{0,-1}^{({\rm a})} \big),\\
&& \!\!\!\!\!\!\!\!\!\!\!\!\!\!\!\!\!\!\!\!
\frac{\partial}{\partial p} \leftrightarrow  -\tfrac{i}{\sqrt{2}}
\big( k \hat{\Delta}_{-1,0}^{({\rm a})} - l \hat{\Delta}_{0,-1}^{({\rm a})} \big),\\
&& \!\!\!\!\!\!\!\!\!\!\!\!\!\!\!\!\!\!\!\! q \leftrightarrow
\tfrac{1}{\sqrt{2}} \big( \hat{\Delta}_{+1,0}^{({\rm a})} \! + \!
\hat{\Delta}_{0,+1}^{({\rm a})} \big) \! - \! \tfrac{1}{2\sqrt{2}}
\big( k \hat{\Delta}_{-1,0}^{({\rm a})} \! + \! l \hat{\Delta}_{0,-1}^{({\rm a})} \big),\\
&& \label{rules-antinorm-4}
\!\!\!\!\!\!\!\!\!\!\!\!\!\!\!\!\!\!\!\! p \leftrightarrow -
\tfrac{i}{\sqrt{2}} \big( \hat{\Delta}_{+1,0}^{({\rm a})} \! - \!
\hat{\Delta}_{0,+1}^{({\rm a})} \big) \! - \! \tfrac{i}{2\sqrt{2}}
\big( k \hat{\Delta}_{-1,0}^{({\rm a})} \! - \! l
\hat{\Delta}_{0,-1}^{({\rm a})} \big),
\end{eqnarray}
\noindent with a displacement operator $\hat{\Delta}_{i,j}^{({\rm
a})}$ for the antinormally ordered moments being
\begin{equation}
\hat{\Delta}_{i,j}^{({\rm a})} \langle \hat{a}^k
(\hat{a}^{\dag})^l \rangle := \langle \hat{a}^{k+i}
(\hat{a}^{\dag})^{l+j} \rangle.
\end{equation}

The time evolution equation for ordered moments is readily
obtained from the Moyal equation
\begin{equation}
\label{Moyal} \!\! \bigg[ \frac{\partial}{\partial t} + p
\frac{\partial}{\partial q} - \sum_{r=0}^{\infty}
\frac{(-1)^r}{(2r+1)!4^r} \frac{d^{2r+1} U}{d q^{2r+1}}
\frac{\partial^{2r+1}}{\partial p^{2r+1}} \bigg] W = 0
\end{equation}
\noindent by substituting normally or antinormally ordered moments
for $W$ and using the correspondence table
(\ref{rules-norm-1})--(\ref{rules-norm-4}) or
(\ref{rules-antinorm-1})--(\ref{rules-antinorm-4}), respectively.

The stationary Schr\"{o}dinger equation, i.e. the eigenstate
problem $\hat{H}|E\rangle = E |E\rangle$, transforms into the
known equation for the Wigner function
\begin{equation}
\label{W-eigenstate} \bigg[ \frac{p^2}{2} - \frac{1}{8}
\frac{\partial^2}{\partial q^2} + \sum_{r=0}^{\infty}
\frac{(-1)^r}{(2r)!4^r} \frac{d^{2r} U}{d q^{2r}}
\frac{\partial^{2r}}{\partial p^{2r}} \bigg] W_E = EW_E
\end{equation}
\noindent and then to the equation for ordered moments according
to associations (\ref{rules-norm-1})--(\ref{rules-norm-4}) and
(\ref{rules-antinorm-1})--(\ref{rules-antinorm-4}).

In order to demonstrate the equations for moments, we consider the
simplest case of the free evolution of the electromagnetic field,
which is effectively governed by the harmonic oscillator potential
$U=q^2/2$. If this is the case, the lowest-order derivatives are
only presented in Eqs. (\ref{Moyal}) and (\ref{W-eigenstate}). For
normally ordered moments we have
\begin{eqnarray}
&& \label{moyal-osc-norm} \!\! \left[ \frac{\partial}{\partial t}
- i (n-m) \right] \langle
(\hat{a}^{\dag})^n \hat{a}^m \rangle =0,\\
&& \label{eigenstate-osc-norm} \!\! \left[
\hat{\Delta}_{+1,+1}^{\rm (n)} + \tfrac{n+m+1}{2} \right] \langle
(\hat{a}^{\dag})^n \hat{a}^m \rangle_E = E \langle
(\hat{a}^{\dag})^n \hat{a}^m \rangle_E.\qquad
\end{eqnarray}
\noindent It is not hard to see that the moments (\ref{Fock-norm})
multiplied by $e^{i(n-m)t}$ are solutions of both Eqs.
(\ref{moyal-osc-norm}) and (\ref{eigenstate-osc-norm}) if
$E=N+1/2$.

Likewise, the antinormally ordered moments (\ref{Fock-antinorm})
multiplied by $e^{i(l-k)t}$ are solutions of equations
\begin{eqnarray}
&& \label{moyal-osc-antinorm} \!\! \left[ \frac{\partial}{\partial
t} + i (k-l) \right] \langle
\hat{a}^k (\hat{a}^{\dag})^l \rangle =0,\\
&& \label{eigenstate-osc-antinorm} \!\! \left[
\hat{\Delta}_{+1,+1}^{\rm (a)} - \tfrac{k+l+1}{2} \right] \langle
\hat{a}^k (\hat{a}^{\dag})^l \rangle_E = E \langle \hat{a}^k
(\hat{a}^{\dag})^l \rangle_E.\qquad
\end{eqnarray}

\section{\label{sec:damping} Damped evolution }
In practice, microwave transmission lines have loss due to a
finite conductivity of waveguide walls or lossy
dielectric~\cite{pozar}. In many experiments the loss may be
neglected. On the other hand, quantum superpositions are very
vulnerable to the relaxation and decoherence while interacting
with the environment. In view of this, the effect of loss is of
interest.

\begin{figure*}
\includegraphics[width=18cm]{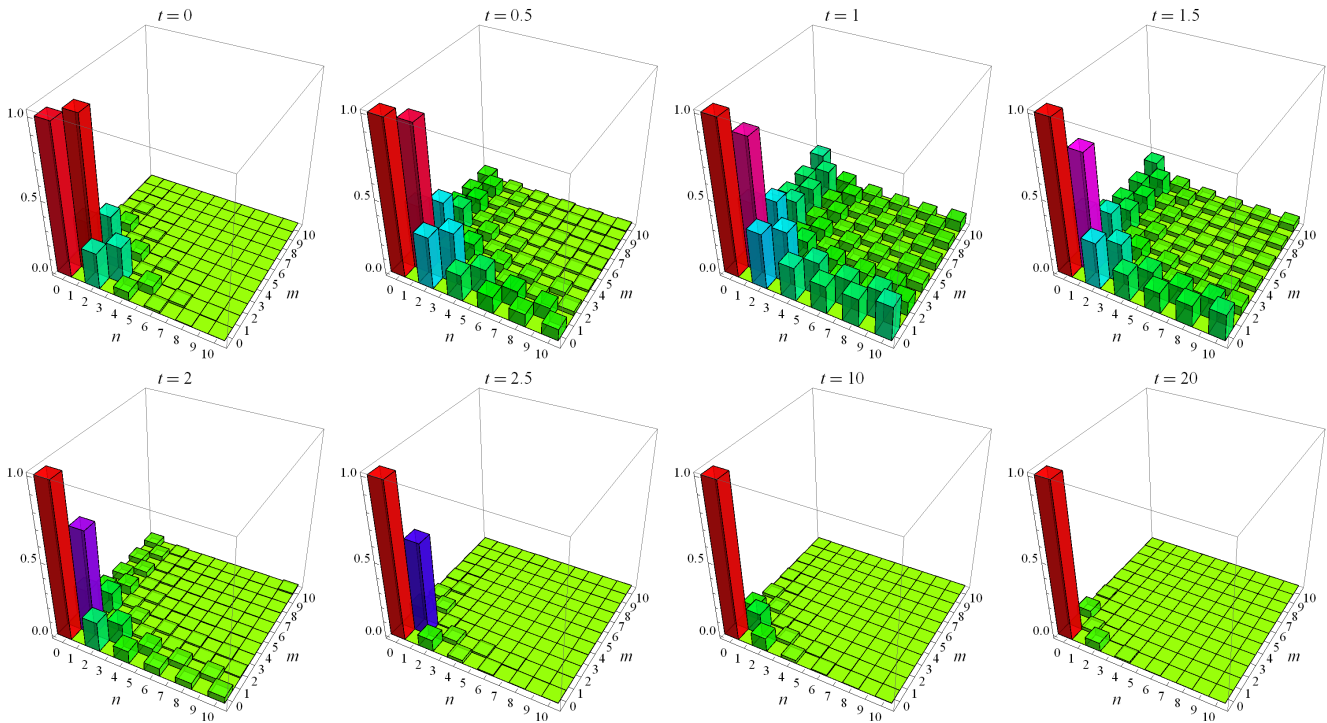}
\caption{\label{figure} (Color online) Snapshots of the normally
ordered moments $| \langle (\hat{a}^{\dag})^n \hat{a}^m \rangle |$
of the odd coherent state (\ref{odd-norm}) with $\alpha=0.5$ at
eight successive times of damped evolution (\ref{f-p-norm}) with
$\gamma=0.1$. }
\end{figure*}

To describe a microwave quantum state in lossy environment we make
use of a damped quantum oscillator
model~\cite{dekker81,dodonov2000}. The Wigner function obeys the
linear Fokker-Planck equation of the form~\cite{dekker81}
\begin{eqnarray}
\label{fokker-planck} && \bigg[ \frac{\partial }{\partial t} + p
\frac{\partial}{\partial
q} - q \frac{\partial}{\partial p} \bigg] W \nonumber\\
&& = \bigg[ 2\gamma \frac{\partial}{\partial p} p +
\frac{\gamma}{2\omega} \left( \frac{\partial^2}{\partial q^2} +
\frac{\partial^2}{\partial p^2} \right) - \frac{\gamma^2}{\omega}
\frac{\partial^2}{\partial p
\partial q} \bigg] W, \qquad
\end{eqnarray}
\noindent where $\gamma$ is the damping coefficient and
$\omega=\sqrt{1-\gamma^2}$. In view of correspondence relations
(\ref{rules-norm-1})--(\ref{rules-norm-4}), Eq.
(\ref{fokker-planck}) transforms into the following equation for
measurable normally ordered moments:
\begin{eqnarray}
\label{f-p-norm} && \left[\frac{\partial}{\partial t} - i(n-m)
\right] \langle (\hat{a}^{\dag})^n \hat{a}^m \rangle = - \gamma \
\Big[ \ n+m
\nonumber\\
&& -n\hat{\Delta}_{-1,+1}^{\rm (n)}-m\hat{\Delta}_{+1,-1}^{\rm
(n)} - (\omega^{-1} - 1) n m \hat{\Delta}_{-1,-1}^{\rm (n)}
\nonumber\\
&& - \textstyle\frac{1}{2} (1+i\omega^{-1}\gamma) n (n-1)
\hat{\Delta}_{-2,0}^{\rm (n)} \nonumber\\
&& - \textstyle\frac{1}{2} (1-i\omega^{-1}\gamma) m (m-1)
\hat{\Delta}_{0,-2}^{\rm (n)} \Big] \langle (\hat{a}^{\dag})^n
\hat{a}^m \rangle.
\end{eqnarray}
\noindent In the same way, one can construct a time evolution
equation for antinormally ordered moments.

The infinite system of equations (\ref{f-p-norm}) is linear. Thus,
the partial differential equation (\ref{fokker-planck}) is reduced
to a linear system of the first order differential equations. This
means that a formal solution of the system is the matrix exponent
which is easy to compute if one is interested in time evolution of
the lower-order moments. Moreover, any linear dynamics of the
system (not necessarily quadratic as in Eq. (\ref{fokker-planck}))
can be predicted numerically in terms of moments, whereas the
dynamics of the Wigner function would require solving
finite-difference equations with artificial mesh spacing.
Conversely, moments $\langle (\hat{a}^{\dag})^n \hat{a}^m \rangle$
are endowed by a natural `mesh spacing' $\Delta n, \Delta m = \pm
1$ and, last but not least, are experimentally accessible.

An example of non-unitary evolution of the normally ordered
moments is presented in Fig.~\ref{figure}. The odd coherent state
$| - \rangle$ (\ref{odd-norm}) with $\alpha=0.5$ evolves in time
according to the system of equations (\ref{f-p-norm}), where
damping coefficient $\gamma=0.1$. Observation of the moments at
successive times provides snapshots of the decoherence process
(cf. \cite{haroche08}).

\section{\label{sec:conclusions} Conclusions }

To resume we point out the main results of our paper.

We have considered two approaches to measuring microwave quantum
states, namely, the homodyne detection scheme with the `optical'
tomogram as output and the heterodyne detection scheme with output
in the form of ordered moments of photon creation and annihilation
operators. A microwave one-mode quantum state can be identified
either with the tomogram $w(X,\theta)$; or the tomographic moments
$\langle X_{\theta}^r \rangle$, $r=0,1,\ldots$; or the normally
ordered moments $\langle (\hat{a}^{\dag})^n \hat{a}^m \rangle$,
$n,m=0,1,\ldots$; or the antinormally ordered moments $\langle
\hat{a}^k (\hat{a}^{\dag})^l \rangle$, $k,l=0,1,\ldots$. It was
shown how to extract these quantities from measurable
characteristics of the amplified microwave signal. We suggest
using the established relations between the tomogram and the
ordered moments
(\ref{tom-norm-moments})--(\ref{norm-mom-tomogram}),
(\ref{antinorm-mom-tomogram}) as well as the relations between the
tomographic moments and the ordered moments
(\ref{norm-mom-tom-mom}), (\ref{antinorm-mom-tom-mom}),
(\ref{tom-mom-norm-mom}) as a cross check of the experimental
results obtained in Ref.~\cite{mallet11} and in
Refs.~\cite{menzel10,mariantoni10,eichler10}.

As the normally/antinormally ordered moments are measurable and
determine a quantum state, an effort to obtain new results for the
ordered moments has been made. Indeed, purity is expressed through
moments and is used in purity-dependent uncertainty relation in
terms of moments. Another result is that the moments are to
satisfy a generalization of the inequality
(\ref{matrix-inequality}). We have obtained the time evolution
equation in terms of moments, which is informationally equivalent
to the von Neumann equation for the density operator and the Moyal
equation for the Wigner function. The energy level problem and the
non-unitary evolution of the damped microwave electromagnetic
field are also considered in terms of moments. The damped
evolution is described by a system of linear differential
equations on moments, which is beneficial (from the viewpoint of
numerical analysis) in comparison with a partial differential
equation on the Wigner function.

Since normally/antinormally ordered moments are of great interest,
a construction of the star-product scheme~\cite{omanko-jpa-02} for
such moments is a problem for further investigation and will be
considered elsewhere.

\begin{acknowledgments}
The authors thank the Russian Foundation for Basic Research for
partial support under Projects Nos. 09-02-00142, 10-02-00312, and
11-02-00456. S.N.F. thanks the Russian Science Support Foundation
for support under Project `Best postgraduates of the Russian
Academy of Sciences 2010', the Dynasty Foundation, and the
Ministry of Education and Science of the Russian Federation for
support under Projects Nos. 2.1.1/5909, $\Pi$558, 14.740.11.0497,
and 14.740.11.1257.
\end{acknowledgments}

\bibliography{Measuring-microwave}

\end{document}